\documentclass[12pt]{article}
\usepackage{amsfonts,amssymb,mathrsfs,hyperref}
\begin{document}
\begin{center}
{\Large\bf Light-cone gauge Hamiltonian for $AdS_4\times\mathbb{CP}^3$ superstring}\\[0.5cm]
{\large D.V.~Uvarov\footnote{E-mail: d\_uvarov@\,hotmail.com, uvarov@\,kipt.kharkov.ua}}\\[0.2cm]
{\it NSC Kharkov Institute of Physics and Technology,}\\ {\it 61108 Kharkov, Ukraine}\\[0.5cm]
\end{center}
\begin{abstract}
It is developed the phase-space formulation for the Type IIA superstring on the $AdS_4\times\mathbb{CP}^3$
background in the $\kappa-$symmetry light-cone gauge for which the light-like
directions are taken from the $D=3$ Minkowski boundary of $AdS_4$.
After fixing bosonic light-cone gauge the superstring Hamiltonian is expressed as a function of the transverse physical variables and in the quadratic approximation corresponds to the light-cone gauge-fixed IIA superstring in flat space.
\end{abstract}

\section{Introduction}
Verification of the $AdS_4/CFT_3$ duality conjecture by Aharony,
Bergman, Jafferis and Maldacena (ABJM) \cite{ABJM} requires on the
string side deeper insights into the structure of the Type IIA
theory on $AdS_4\times\mathbb{CP}^3$ background. Initial proposal
\cite{AF}, \cite{Stefanski}\footnote{Another way to construct the
$AdS_4\times\mathbb{CP}^3$ superstring based on the pure spinor
approach was pursued in \cite{PS}.} to construct the Green-Schwarz
superstring action on such a background using the supercoset
approach \cite{MT}, \cite{Kallosh} was based on the observation
that the isometry groups $SO(2,3)\sim Sp(4)$ and $SU(4)\sim SO(6)$
of the $AdS_4=SO(2,3)/SO(1,3)$ and $\mathbb{CP}^3=SU(4)/U(3)$
parts of the background match the bosonic subgroup $Sp(4)\times
SO(6)$ of the $OSp(4|6)$ supergroup. The resulting
$OSp(4|6)/(SO(1,3)\times U(3))$ supercoset action \cite{AF},
\cite{Stefanski} is the functional of 10 bosonic and 24 fermionic
coordinate fields of the 'reduced' target superspace invariant
under the $OSp(4|6)$ isometry supergroup of the
$AdS_4\times\mathbb{CP}^3$ background. It is by construction
classically integrable and invariant under the $8-$parameter
$\kappa-$symmetry transformations. The $OSp(4|6)/(SO(1,3)\times
U(3))$ supercoset model corresponds to the full Type IIA
$AdS_4\times\mathbb{CP}^3$ superstring action \cite{GSWnew}, in
which the $\kappa-$symmetry has been partially fixed by putting to
zero those coordinates associated with 8 broken space-time
supersymmetries\footnote{Such full action does not correspond to a
$2d$ supercoset sigma-model but can be obtained by performing the
double dimensional reduction \cite{DHIS} of the $D=11$
supermembrane on the maximally supersymmetric $AdS_4\times S^7$
background, whose action can be constructed within the
$OSp(4|8)/(SO(1,3)\times SO(7))$ supercoset approach
\cite{Plefka98}, provided the 7-sphere is realized in the Hopf
fibration form as $\mathbb{CP}^3\times U(1)$ \cite{Pope},
\cite{Volkov}.}. Besides that such gauge choice narrows down the
possibilities of further gauge fixing and hence the action
simplification within the $OSp(4|6)/(SO(1,3)\times U(3))$
supercoset model it has the limited range of validity restricted
to the string configurations propagating in $\mathbb{CP}^3$ or
both parts of the background \cite{AF}\footnote{String
configurations lying solely in the $AdS$ part of the background
are of interest, e.g., for studying the string quantization around
particular classical solutions \cite{GKP}, \cite{FT}. Related
issues for the ABJM correspondence have been addressed in
\cite{semiclas}.}.

This motivates examining gauge conditions other than that used to
obtain the $OSp(4|6)/(SO(1,3)\times U(3))$ supercoset action. In
particular, in Ref. \cite{me-lc} we have considered the fermionic
light-cone gauge for which both light-like directions are on the
$D=3$ Minkowski boundary of the $AdS_4$ space and there are set to
zero the fermionic coordinates positively charged w.r.t. $SO(1,1)$
light-cone isometry group. The resulting gauge-fixed Lagrangian
includes fermions up to the 4th power and manifestly exhibits the
$SU(3)$ symmetry subgroup of the $SU(4)$ global symmetry of
$\mathbb{CP}^3$. To make contact with the $3d$ extended
superconformal symmetry that is the symmetry group on the gauge
theory side of ABJM duality we have associated the IIA superspace
coordinates with the generators of the $AdS_4\times S^7$ isometry
supergroup $OSp(4|8)$ that is broken down to $OSp(4|6)$ by the
$AdS_4\times\mathbb{CP}^3$ background and elaborated on the
$osp(4|8)$ and $osp(4|6)$ algebras realization as extended $D=3$
superconformal algebras \cite{U}. Such a realization of the
$AdS_4\times\mathbb{CP}^3$ background isometry superalgebra
implies splitting fermionic coordinates into those corresponding
to super-Poincare (linearly realized) and superconformal
(non-linearly realized) symmetries\footnote{The splitting of the
fermionic generators/coordinates into the
supersymmetry/superconformal ones in the sense of the CFT on the
$AdS$ boundary was introduced in \cite{9807115}-\cite{PST} for the
study of brane models on the $AdS\times S$ backgrounds.}.

In the present paper we study the Hamiltonian formulation for the
fermionic light-cone gauge Lagrangian derived in \cite{me-lc}, the
Hamiltonian framework being more systematic for treating
constrained dynamical systems as is the $AdS_4\times\mathbb{CP}^3$
superstring. Besides that bosonic light-cone gauge for the curved
background under consideration can be fixed within the Hamiltonian
approach quite similarly to the flat-space and $AdS_5\times S^5$
cases \cite{GGRT}-\cite{uniform-lc}, whereas conventional in the
Lagrangian approach conformal gauge condition on the $2d$ metric
appears to be incompatible with the bosonic light-cone gauge
$x^+=\tau$ \footnote{The discussion of the issues related to
bosonic light-cone gauge fixing for strings on the $AdS$
background can be found in \cite{MTlc}.}. Our main result is the
derivation of the light-cone gauge Hamiltonian as a function of
8+8 physical variables. Being quartic in the fermions it still
appears to be highly nonlinear in the bosonic fields.

In the quadratic approximation that is the leading one in the
large tension (light-cone momentum) limit obtained Hamiltonian
reduces to the light-cone gauge IIA superstring Hamiltonian in
flat space \cite{GSW} and thus is different from that for the
$AdS_4\times\mathbb{CP}^3$ $pp-$wave superstring \cite{Sugiyama},
\cite{Nishioka} corresponding to the leading order approximation
for the light-cone gauge choice of \cite{Astolfi}-\cite{Zarembo}.
This is because of our $\kappa-$symmetry gauge choice that amounts
to picking the null geodesic from the $AdS_4$ part of the
background. As is known (see, e.g., \cite{BFFP}) for the
$AdS\times S$ type backgrounds, as well as for
$AdS_4\times\mathbb{CP}^3$ there exist two distinct types of null
geodesics. Whenever null geodesic lies entirely within the $AdS$
subspace, as in our case, the resulting $pp-$wave is necessarily
flat. The nontrivial $pp-$wave can be obtained by choosing the
tangent vector to the geodesic to have nonzero components in the
directions tangent to the compact space $S$ or $\mathbb{CP}^3$. In
both cases the light-cone gauge $pp-$wave superstring action
corresponds to the $2d$ free-field theory and can be explicitly
quantized \cite{GSW}, \cite{MT-ppwave}, \cite{Sugiyama}. Further
terms in the Lagrangian/Hamiltonian perturbative expansion can be
arranged as the power series with the their geometric
interpretation provided in \cite{Blau}.

\section{Hamiltonian for the light-cone gauge $AdS_4\times\mathbb{CP}^3$ superstring}

The $AdS_4$ part of the background in the Poincare patch can be
parametrized by the coordinates $(x^\pm=x^2\pm x^0,\ x^1,\
\varphi)$, where $x^m=(x^0, x^1, x^2)$ are $3d$ Minkowski boundary
coordinates, while $\mathbb{CP}^3$ is parametrized by the
coordinates $z^M$, $M=1,...,6$ \footnote{We have left unspecified
at this stage the explicit $\mathbb{CP}^3$ parametrization and
metric.}.

We consider the Grassmann-odd coordinates $\theta^\mu_{A'}$ as associated with the $D=3$ $\mathcal N=8$ super-Poincare generators of $osp(4|8)$ and $\eta_{\mu A'}$ - with the superconformal generators. As it follows from the $AdS_4\times S^7$ supermembrane consideration \cite{Plefka98} they have the transformation properties of the $D=3$ (in index $\mu$) and $D=7$ (in index $A'$) Majorana spinors and can be split as follows
\begin{equation}
\theta^\mu_{A'}=(
\theta^\mu_A,\
\bar\theta^{\mu A}),\quad
\eta_{\mu A'}=(
\eta_{\mu A},\
\bar\eta_\mu^A),
\end{equation}
according to the $SU(4)$ decomposition of the minimal spinor representation of $Spin(7)$ as $\mathbf{8}=\mathbf{4}\oplus\mathbf{\bar 4}$. Using further the decomposition of $\mathbf{4}$ and $\mathbf{\bar 4}$ representations into the $SU(3)$ irreducible ones as $\mathbf{4}=\mathbf{3}\oplus\mathbf{1}$, $\mathbf{\bar 4}=\mathbf{\bar 3}\oplus\mathbf{1}$ we obtain
\begin{equation}
\theta^\mu_A=(\theta^\mu_a, \theta^\mu_4),\quad\bar\theta^{\mu A}=(\bar\theta^{\mu a}, \bar\theta^{\mu 4}),\quad
\eta_{\mu A}=(\eta_{\mu a}, \eta_{\mu 4}),\quad\bar\eta_\mu^A=(\bar\eta_\mu^a, \bar\eta_\mu^4).
\end{equation}
The coordinates $(\theta^\mu_a, \bar\theta^{\mu a})$ and $(\eta_{\mu a},\bar\eta_\mu^a)$ transforming in the $\mathbf{3}$ and $\mathbf{\bar 3}$ representations of $SU(3)$ are associated with the 24 unbroken Type IIA supersymmetries viewed as the $D=3$ $\mathcal N=6$ super-Poincare and superconformal symmetries from the $AdS_4$ boundary perspective, other 8 coordinates $(\theta_4, \bar\theta^4)$ and $(\eta_4,\bar\eta^4)$ correspond to the  supersymmetries broken by the $AdS_4\times\mathbb{CP}^3$ background.

In \cite{me-lc} there was considered the fermionic light-cone gauge condition
\begin{equation}
\mathfrak{g}^{+2}{}_{\hat\alpha}{}^{\hat\beta}\Theta_{\hat\beta}
=(\mathfrak{g}^{0}+\mathfrak{g}^2)_{\hat\alpha}{}^{\hat\beta}\Theta_{\hat\beta}=0,
\end{equation}
where $32\times32$ $\gamma-$matrices $\mathfrak{g}^{\hat m'}{}_{\hat\alpha}{}^{\hat\beta}$ have been defined in Appendix A of Ref.~\cite{me-lc} and $\Theta_{\hat\beta}=(\theta_{\nu B}, \eta^\nu_B, \bar\theta^B_\nu, \bar\eta^{\nu B})$, that in components reads
\begin{equation}\label{lcfermi}
\theta^2_A=\bar\theta^{2 A}=\eta_{1 A}=\bar\eta_{1}^{A}=0.
\end{equation}
It is characterized by setting to zero odd coordinates associated
with the generators of $osp(4|8)$ negatively charged w.r.t. the
$so(1,1)$ generator $M^{+-}\equiv2M^{02}$ from the $AdS_4$
boundary Lorentz group. Remaining odd coordinates
\begin{equation}
\theta^1_A=\theta^-_A=\theta_A,\quad\bar\theta^{1A}=\bar\theta^{-A}=\bar\theta^A,\quad
\eta^{1}_{A}=\eta^{-}_{A}=\eta_A,\quad\bar\eta^{1A}=\bar\eta^{-A}=\bar\eta^A
\end{equation}
become physical fermionic fields of the $AdS_4\times\mathbb{CP}^3$
superstring in the gauge (\ref{lcfermi})\footnote{Observe that
among the physical fermions there are those associated with the
unbroken, as well as the broken space-time supersymmetries.}.

The fermionic light-cone gauge superstring action obtained in \cite{me-lc} can be presented as
\begin{equation}
\mathscr S=\int d\tau d\sigma\mathscr L(\tau,\sigma),
\end{equation}
where the corresponding Lagrangian density is given by
\begin{equation}\label{initiall}
\begin{array}{rl}
\mathscr L=&-\frac{T}{2}\gamma^{ij}\left\{\frac{e^{-4\varphi}}{4}\left[\frac12(\partial_ix^+\partial_jx^-+\partial_jx^+\partial_ix^-)
+\partial_ix^1\partial_j x^1\right]+\partial_i\varphi\partial_j\varphi+g_{MN}\partial_iz^M\partial_jz^N\right.\\[0.2cm]
+&\left.\frac{e^{-2\varphi}}{4}(\partial_ix^+\varpi_j+\partial_jx^+\varpi_i)
+(\partial_ix^+\partial_jz^M+\partial_jx^+\partial_iz^M)q_M+B\partial_ix^+\partial_jx^+\right\}\\[0.2cm]
+&T\varepsilon^{ij}\left(\tilde\omega_i\partial_jx^++C\partial_ix^1\partial_jx^++\partial_ix^+\partial_jz^M\tilde q_M\right).
\end{array}
\end{equation}
In the Lagrangian (\ref{initiall}) we have introduced the following quantities
\begin{equation}
\begin{array}{rl}
\varpi_i=&ie^{-2\varphi}(\partial_i\theta_a\bar\theta^a-\theta_a\partial_i\bar\theta^a)
+i(\partial_i\theta_4\bar\theta^4-\theta_4\partial_i\bar\theta^4)
+ie^{2\varphi}(\partial_i\eta_a\bar\eta^a-\eta_a\partial_i\bar\eta^a)\\[0.2cm]
+&i(\partial_i\eta_4\bar\eta^4-\eta_4\partial_i\bar\eta^4),\\[0.2cm]
\tilde\omega_i=&\frac{e^{-2\varphi}}{2}(\hat\eta_a\hat\partial_i\bar\theta^a+\hat\partial_i\theta_a\hat{\bar\eta}{}^a)
+\frac{e^{-2\varphi}}{4}(\partial_i\theta_4\bar\eta^4-\partial_i\eta_4\bar\theta^4
+\eta_4\partial_i\bar\theta^4-\theta_4\partial_i\bar\eta^4),\\[0.2cm]
B=&2[(\hat\eta_a\hat{\bar\eta}{}^a)^2+e^{-\varphi}(\varepsilon_{abc}\hat{\bar\eta}{}^a\hat{\bar\eta}{}^b\hat{\bar\eta}{}^c\bar\eta^4+\varepsilon^{abc}\hat\eta_a\hat\eta_b\hat\eta_c\eta_4)+2e^{-2\varphi}\eta_4\bar\eta^4(\hat\eta_a\hat{\bar\eta}{}^a-
e^{-2\varphi}\theta_4\bar{\theta}{}^4)],\\[0.2cm] 
C=&e^{-2\varphi}(\hat\eta_a\hat{\bar\eta}{}^a+\frac{e^{-2\varphi}}{2}\Theta),\ \Theta=\theta_4\bar{\theta}{}^4+\eta_4\bar{\eta}{}^4,\\[0.2cm]
q_M=&\frac12(\Omega^a_M\varepsilon_{abc}\hat{\bar\eta}{}^b\hat{\bar\eta}{}^c-\Omega_{aM}\varepsilon^{abc}\hat\eta_b\hat\eta_c)+e^{-\varphi}(\Omega_{aM}\hat{\bar\eta}{}^a\bar{\eta}{}^4-\Omega^a_M\hat{\eta}_a\eta_4)
+e^{-2\varphi}\Theta\tilde\Omega_a{}^a{}_M,\\[0.2cm]
\tilde
q_M=&ie^{-\varphi}\left[\Omega_{aM}\hat{\bar\eta}{}^a\bar\theta^4+\Omega^a_M\hat\eta_a\theta_4
+e^{-\varphi}(\theta_4\bar{\eta}{}^4-\eta_4\bar{\theta}{}^4)\tilde\Omega_a{}^a{}_M\right].
\end{array}
\end{equation}
The 1-forms $\Omega_a(d)=\Omega_{aM}dz^M$,
$\Omega^a(d)=\Omega^a{}_Mdz^M$, being the $su(4)/u(3)$ Cartan
forms, define the $\mathbb{CP}^3$ complex vielbein such that
$ds^2_{CP^3}=\Omega_a(d)\Omega^a(d)$, while
$\tilde\Omega_a{}^a(d)$ enters the background RR 1-form potential
superfield \cite{me-lc}, \cite{U}. The Grassmann variables with
hats differ from those without in the multiplication by the matrix
$T_{\hat a}{}^{\hat b}$ \cite{me-lc}, \cite{U}
\begin{equation}\label{matrixt}
T_{\hat a}{}^{\hat b}(z)=\left(
\begin{array}{cc}
T_a{}^b & T_{ab}\\[0.2cm]
T^{ab} & T^a{}_b
\end{array}\right)
\end{equation}
as
\begin{equation}
\hat d\theta_a(\hat\eta_a)=T_a{}^bd\theta_b(\eta_b)+T_{ab}d\bar\theta^b(\bar\eta^b),\quad \hat
d\bar\theta^a(\hat{\bar\eta}{}^a)=T^a{}_bd\bar\theta^b(\bar\eta^b)+T^{ab}d\theta_b(\eta_b).
\end{equation}

The string tension induced as a result of the dimensional reduction of the $AdS_4\times S^7$ supermembrane is $T=\frac{R^2}{k^2l^4_{Pl}}$, where $R$ is the $S^7$ radius, $l_{Pl}$ is the $11d$ Planck length, and $k$ is the Chern-Simons level from the boundary ABJM theory perspective. If the bosonic coordinates are measured in the units of $(Rl_{Pl}k)^{1/2}$ and fermionic in the units of $(Rl_{Pl}k)^{1/4}$ then the dimensionless tension becomes $\frac{1}{k}(\frac{R}{l_P})^3$ and is suggested to be equal to $2^{5/2}\pi\sqrt{\lambda}$, where $\lambda$ is the 't Hooft coupling in the ABJM theory \cite{ABJM}.

From the Lagrangian (\ref{initiall}) we deduce the momenta densities conjugate to the coordinates from the $AdS_4$
\begin{equation}
\begin{array}{rl}
p_-(\tau,\sigma)=-&\frac{Te^{-4\varphi}}{8}\gamma^{\tau i}\partial_i x^+,\\[0.2cm]
p_+(\tau,\sigma)=-&T\gamma^{\tau i}\left(\frac{e^{-4\varphi}}{8}\partial_i x^-+B\partial_i x^++q_M\partial_iz^M+\frac{e^{-2\varphi}}{4}\varpi_i\right)\\[0.2cm]
-&T(C\partial_\sigma x^1-\tilde q_M\partial_\sigma z^M+\tilde\omega_\sigma),\\[0.2cm]
p_1(\tau,\sigma)=-&\frac{Te^{-4\varphi}}{4}\gamma^{\tau i}\partial_i x^1+TC\partial_\sigma x^+,\\[0.2cm]
p_\varphi(\tau,\sigma)=-&T\gamma^{\tau i}\partial_i\varphi
\end{array}
\end{equation}
and $\mathbb{CP}^3$ parts of the background
\begin{equation}
p_M(\tau,\sigma)=-T\gamma^{\tau i}(g_{MN}\partial_iz^N+q_M\partial_i x^+)-T\tilde q_M\partial_\sigma x^+.
\end{equation}
From these relations one can express the world-sheet time derivatives of the bosonic fields
\begin{equation}\label{velocities}
\begin{array}{rl}
\partial_\tau x^+=&-\frac{8e^{4\varphi}}{T\gamma^{\tau\tau}}p_-
-\frac{\gamma^{\tau\sigma}}{\gamma^{\tau\tau}}\partial_\sigma x^+,
\\[0.2cm]
\partial_\tau x^-=&-2e^{2\varphi}\varpi_\tau+\frac{1}{\gamma^{\tau\tau}}A^-
-\frac{\gamma^{\tau\sigma}}{\gamma^{\tau\tau}}(\partial_\sigma x^-+2e^{2\varphi}\varpi_\sigma),\\[0.2cm]
\partial_\tau x^1=&-\frac{4e^{4\varphi}}{T\gamma^{\tau\tau}}(p_1-TC\partial_\sigma x^+)-\frac{\gamma^{\tau\sigma}}{\gamma^{\tau\tau}}\partial_\sigma x^1,\\[0.2cm]
\partial_\tau \varphi=&-\frac{1}{T\gamma^{\tau\tau}}p_\varphi-\frac{\gamma^{\tau\sigma}}{\gamma^{\tau\tau}}\partial_\sigma \varphi,\\[0.2cm]
\partial_\tau z^M=&-\frac{g^{MN}}{T\gamma^{\tau\tau}}(p_N-8e^{4\varphi}p_-q_N+T\tilde q_N\partial_\sigma x^+)-\frac{\gamma^{\tau\sigma}}{\gamma^{\tau\tau}}\partial_\sigma z^M,
\end{array}
\end{equation}
where
\begin{equation}
A^-=-\textstyle{\frac{8e^{4\varphi}}{T}}\left(p_+-8e^{4\varphi}p_-(B-(q\cdot q))-(p\cdot q)-T(q\cdot\tilde q)\partial_\sigma x^++TC\partial_\sigma x^1-T\tilde q_M\partial_\sigma z^M+T\tilde\omega_\sigma\right)
\end{equation}
and the following notation for the scalar product with the inverse $\mathbb{CP}^3$ metric $g^{MN}(z)$ has been introduced $q_Mg^{MN}q_N=(q\cdot q)$ etc.
Substituting (\ref{velocities}) back into (\ref{initiall}) we are able to obtain the phase-space representation for the fermionic light-cone gauge fixed
Lagrangian of the $AdS_4\times\mathbb{CP}^3$ superstring
\begin{equation}\label{initiall1storder}
\mathscr L=p_+\partial_\tau x^++p_-\partial_\tau
x^-+p_1\partial_\tau x^1+p_\varphi\partial_\tau
\varphi+p_M\partial_\tau
z^M+2e^{2\varphi}p_-\varpi_\tau+T\tilde\omega_\tau\partial_\sigma
x^++{\textstyle\frac{1}{\gamma^{\tau\tau}}}T_1+{\textstyle\frac{\gamma^{\tau\sigma}}{\gamma^{\tau\tau}}}T_2.
\end{equation}
The last two summands introduce via the Lagrange multipliers $\frac{1}{\gamma^{\tau\tau}}$ and $\frac{\gamma^{\tau\sigma}}{\gamma^{\tau\tau}}$ the Virasoro constraints expressed in terms of the phase-space variables
\begin{equation}
\begin{array}{rl}
T_1=&\frac{8e^{4\varphi}}{T}p_+p_-+\frac{Te^{-4\varphi}}{8}\partial_\sigma x^+\partial_\sigma x^-+\frac{2e^{4\varphi}}{T}p^2_1
+\frac{Te^{-4\varphi}}{8}\partial_\sigma x^1\partial_\sigma x^1+\frac{1}{2T}p^2_\varphi+\frac{T}{2}\partial_\sigma\varphi\partial_\sigma\varphi\\[0.2cm]
+&\frac{1}{2T}(p\cdot p)+\frac{T}{2}\partial_\sigma z^Mg_{MN}\partial_\sigma z^N+\frac{32e^{8\varphi}}{T}p^2_-((q\cdot q)-B)\\[0.2cm]
+&8e^{4\varphi}p_-(\tilde\omega_\sigma-\frac{1}{T}(p\cdot q)-(q\cdot\tilde q)\partial_\sigma x^++C\partial_\sigma x^1-\tilde q_M\partial_\sigma z^M)\\[0.2cm]
-&(\frac{Te^{-2\varphi}}{2}\varpi_\sigma-Tq_M\partial_\sigma z^M-(p\cdot\tilde q)+4e^{4\varphi}p_1C)\partial_\sigma x^+\\[0.2cm]
+&\frac{T}{2}((\tilde q\cdot\tilde q)+B+4e^{4\varphi}C^2)\partial_\sigma x^+\partial_\sigma x^+\approx0,
\end{array}
\end{equation}
\begin{equation}
T_2=p_+\partial_\sigma x^++p_-\partial_\sigma
x^-+p_1\partial_\sigma x^1+p_\varphi\partial_\sigma
\varphi+p_M\partial_\sigma
z^M+2e^{2\varphi}p_-\varpi_\sigma+T\tilde\omega_\sigma\partial_\sigma
x^+\approx0.
\end{equation}

The bosonic light-cone gauge conditions we consider are analogous
to those of Refs.\cite{GGRT}, \cite{MTlc}
\begin{equation}
x^+(\tau,\sigma)=\tau,\quad p_-(\tau,\sigma)=\frac{\tilde
P_-}{2}=const.
\end{equation}
Using the equation of motion for $p_+$ it is possible to express $\gamma^{\tau\tau}$ through $\tilde P_-$
\begin{equation}
\gamma^{\tau\tau}=-\frac{4e^{4\varphi}}{T}\tilde P_-.
\end{equation}
Further in order bring the symplectic structure for the fermions
defined by the $2e^{2\varphi}p_-\varpi_\tau$ term in
(\ref{initiall1storder}) to the simplest form the following
rescaling is performed
\begin{equation}\label{rescale-f}
\begin{array}{c}
\theta_a(\bar\theta^a)\to\frac{1}{\sqrt{\tilde P_-}}\theta_a(\bar\theta^a);\quad\eta_a(\bar\eta^a)\to \frac{\exp\left(-2\varphi/{\scriptstyle\sqrt{\tilde P_-}}\right)}{\sqrt{\tilde P_-}}\eta_a(\bar\eta^a);\\[0.2cm]
\theta_4(\bar\theta^4)\to\frac{\exp\left(-\varphi/{\scriptstyle\sqrt{\tilde P_-}}\right)}{\sqrt{\tilde P_-}}\theta_4(\bar\theta^4);\quad\eta_4(\bar\eta^4)\to\frac{\exp\left(-\varphi/{\scriptstyle\sqrt{\tilde P_-}}\right)}{\sqrt{\tilde P_-}}\eta_4(\bar\eta^4)
\end{array}
\end{equation}
accompanied by the canonical rescaling of the transverse bosonic phase-space variables
\begin{equation}\label{rescale-b}
\begin{array}{c}
x^1\to\frac{1}{\sqrt{\tilde P_-}}x^1,\quad\varphi\to\frac{1}{\sqrt{\tilde P_-}}\varphi,\quad z^M\to\frac{1}{\sqrt{\tilde P_-}}z^M;\\[0.2cm]
p_1\to{\scriptstyle\sqrt{\tilde P_-}}p_1,\quad p_\varphi\to{\scriptstyle\sqrt{\tilde P_-}}p_\varphi,\quad p_M\to{\scriptstyle\sqrt{\tilde P_-}}p_M
\end{array}
\end{equation}
that will allow to perform the power series expansion of the Hamiltonian in the large $\tilde P_-$ limit.

The light-cone gauge fixed Lagrangian (\ref{initiall1storder}) then takes the form
\begin{equation}\label{boslc}
\begin{array}{rl}
\mathscr L_{l.c.}=&p_1\partial_\tau x^1+p_\varphi\partial_\tau\varphi+p_M\partial_\tau z^M
+i(\partial_\tau\theta_A\bar\theta^A-\theta_A\partial_\tau\bar\theta^A\\[0.2cm]
+&\partial_\tau\eta_A\bar\eta^A-\eta_A\partial_\tau\bar\eta^A)-\mathscr H_{l.c.}.
\end{array}
\end{equation}
Corresponding light-cone gauge Hamiltonian density is given by the utmost quartic in the Grassmann coordinates expression
\begin{equation}\label{hlc}
\begin{array}{rl}
\mathscr H_{l.c.}=&\frac{\tilde
T}{2}\exp\left(-4\varphi/{\scriptstyle\sqrt{\tilde
P_-}}\right)\left[\hat\eta_a\hat\partial_\sigma\bar\theta^a\!+\!\hat\partial_\sigma\theta_a\hat{\bar\eta}{}^a
\!+\!\frac12(\partial_\sigma\theta_4\bar\eta^4\!-\!\partial_\sigma\eta_4\bar\theta^4\!+\!
\eta_4\partial_\sigma\bar\theta^4\!-\!\theta_4\partial_\sigma\bar\eta^4)\right]\\[0.2cm]
+&\frac{\tilde T}{4}\exp\left(-4\varphi/{\scriptstyle\sqrt{\tilde
P_-}}\right)\left[\frac{2}{\tilde
T}\exp\left(4\varphi/{\scriptstyle\sqrt{\tilde P_-}}\right)p^2_1
+\frac{\tilde T}{8}\exp\left(-4\varphi/{\scriptstyle\sqrt{\tilde P_-}}\right)\partial_\sigma x^1\partial_\sigma x^1\right.\\[0.2cm]
+&\frac{1}{2\tilde T}p^2_\varphi+\frac{\tilde T}{2}\partial_\sigma\varphi\partial_\sigma\varphi+\frac{1}{2\tilde T}(p\cdot p)+\frac{\tilde T}{2}\partial_\sigma z^Mg_{MN}\partial_\sigma z^N\\[0.2cm]
+&\frac{8\tilde P_-}{\tilde
T}\exp\left(8\varphi/{\scriptstyle\sqrt{\tilde
P_-}}\right)((q\cdot q)-B)
+4\exp\left(4\varphi/{\scriptstyle\sqrt{\tilde P_-}}\right)\sqrt{\tilde P_-}C\partial_\sigma x^1\\[0.2cm] -&\left.4\exp\left(
4\varphi/{\scriptstyle\sqrt{\tilde P_-}}\right)\sqrt{\tilde
P_-}\tilde q_M\partial_\sigma z^M-\frac{4}{\tilde
T}\exp\left(4\varphi/{\scriptstyle\sqrt{\tilde
P_-}}\right)\sqrt{\tilde P_-}(p\cdot q)\right],
\end{array}
\end{equation}
where $\tilde T=\frac{T}{\tilde P_-}$.

As usual in the light-cone gauge approach the $T_2\approx0$ constraint can be explicitly solved for $\partial_\sigma x^-$ that decouples from the Hamiltonian. Its only nontrivial content is the zero-mode part that defines the phase-space representation of the level matching condition
\begin{equation}
{\textstyle\frac{1}{\tilde P_-}\oint} d\sigma[p_1\partial_\sigma
x^1+p_\varphi\partial_\sigma \varphi+p_M\partial_\sigma
z^M+i(\partial_\sigma\theta_A\bar\theta^A-\theta_A\partial_\sigma\bar\theta^A+
\partial_\sigma\eta_A\bar\eta^A-\eta_A\partial_\sigma\bar\eta^A)]=0.
\end{equation}

Being highly non-linear the light-cone gauge Hamiltonian (\ref{hlc}) can be studied in various simplifying limits (see \cite{found} for a review). The rescalings of the phase-space variables performed in (\ref{rescale-f}), (\ref{rescale-b}) make the form of the Hamiltonian suitable for performing the large $\tilde P_-$ (or equivalently large $T$) expansion with $\tilde T$ fixed that amounts to the power series in the phase-space fields. To this end we choose to consider the $\mathbb{CP}^3$ parametrization by 3 complex coordinates $(z^a, \bar z_a)$ transforming in $\mathbf{3}$ and $\mathbf{\bar 3}$ of $SU(3)$ \cite{U} with the line element
\begin{equation}\label{metric}
ds^2_{CP^3}=g_{ab}dz^adz^b+g^{ab}d\bar z_ad\bar z_b+2g_a{}^bdz^ad\bar z_b.
\end{equation}
The metric components are given by
\begin{equation}
\begin{array}{c}
g_{ab}=\frac{1}{4|z|^4}(|z|^2-\sin^2{|z|}+\sin^4{|z|})\bar z_a\bar z_b,\quad g^{ab}=\frac{1}{4|z|^4}(|z|^2-\sin^2{|z|}+\sin^4{|z|})z^a z^b,\\[0.2cm]
g_a{}^b=\frac{\sin^2{|z|}}{2|z|^2}\delta_a^b+\frac{1}{4|z|^4}(|z|^2-\sin^2{|z|}-\sin^4{|z|})\bar z_az^b,\quad |z|^2=z^a\bar z_a.
\end{array}
\end{equation}

The metric (\ref{metric}) can be related to the conventional Fubini-Study one in the following way. Separate out the norm of $z^a$
\begin{equation}
z^a=|z|u^a,\quad\bar z_a=|z|\bar u_a:\quad u^a\bar u_a=1
\end{equation}
so that (\ref{metric}) acquires the form
\begin{equation}\label{metric2}
ds^2_{CP^3}=d|z|^2+\sin^2{|z|}du^ad\bar u_a-\sin^4{|z|}u^ad\bar u_adu^b\bar u_b.
\end{equation}
Analogously the $\mathbb{CP}^3$ line element for the Fubini-Study metric
\begin{equation}
ds^2_{CP^3}=\frac{1}{1+|w|^2}dw^ad\bar w_a-\frac{1}{(1+|w|^2)^2}w^ad\bar w_adw^b\bar w_b
\end{equation}
after separating the norm in the $w-$coordinates
\begin{equation}
w^a=|w|u^a,\quad\bar w_a=|w|\bar u_a:\quad u^a\bar u_a=1
\end{equation}
is brought to the form
\begin{equation}\label{fubini}
ds^2_{CP^3}=\frac{1}{(1+|w|^2)^2}d|w|^2+\frac{|w|^2}{(1+|w|^2)}du^ad\bar u_a-\frac{|w|^4}{(1+|w|^2)^2}u^ad\bar u_adu^b\bar u_b.
\end{equation}
Comparing (\ref{metric2}) with (\ref{fubini}) we observe that
\begin{equation}
\sin^2{|z|}=\frac{|w|^2}{1+|w|^2}
\end{equation}
and
\begin{equation}
z^a=\frac{\arcsin{\frac{|w|}{\sqrt{1+|w|^2}}}}{|w|}w^a,\quad\bar z_a=\frac{\arcsin{\frac{|w|}{\sqrt{1+|w|^2}}}}{|w|}\bar w_a .
\end{equation}

The components of the metric inverse to (\ref{metric}) read
\begin{equation}
\begin{array}{c}
g^{-1}_{ab}=\frac{|z|^2-\sin^2{|z|}+\sin^4{|z|}}{|z|^2(\sin^4{|z|}-\sin^2{|z|})}\bar z_a\bar z_b,\quad
g^{-1ab}=\frac{|z|^2-\sin^2{|z|}+\sin^4{|z|}}{|z|^2(\sin^4{|z|}-\sin^2{|z|})}z^a z^b,\\[0.2cm]
g^{-1}{}_a{}^b=\frac{2|z|^2}{\sin^2{|z|}}\delta_a^b
+\frac{\sin^2{|z|}(1+2|z|^2)-\sin^4{|z|}-|z|^2}{|z|^2(\sin^2{|z|}-\sin^4{|z|})}\bar z_a z^b.
\end{array}
\end{equation}
Then the metric tensor (\ref{metric}) and its inverse can be
Taylor expanded with the first terms given by
\begin{equation}
\begin{array}{c}
g_{ab}=\left(\frac13-\frac{8}{45}|z|^2\right)\bar z_a\bar z_b+O(6),\quad g^{ab}=\left(\frac13-\frac{8}{45}|z|^2\right)z^a z^b+O(6),\\[0.2cm]
g_a{}^b=\left(\frac12-\frac{1}{6}|z|^2+\frac{1}{45}|z|^4\right)\delta_a^b-\bar z_a z^b\left(\frac16-\frac{7}{45}|z|^2\right)+O(6)
\end{array}
\end{equation}
and
\begin{equation}
\begin{array}{c}
g^{-1ab}=-\left(\frac43+\frac{48}{45}|z|^2\right)z^a z^b+O(6),\quad g^{-1}_{ab}=-\left(\frac43+\frac{48}{45}|z|^2\right)\bar z_a\bar z_b+O(6),\\[0.2cm]
g^{-1}{}_a{}^b=2\left(1+\frac{1}{3}|z|^2+\frac{1}{15}|z|^4\right)\delta_a^b+2\bar
z_a z^b\left(\frac13+\frac{21}{45}|z|^2\right)+O(6).
\end{array}
\end{equation}
For the matrix $T_{\hat a}{}^{\hat b}$ (\ref{matrixt}) that for the chosen $\mathbb{CP}^3$ parametrization takes the form \cite{U}
\begin{equation}
T_{\hat a}{}^{\hat b}=\left(
\begin{array}{cc}
\delta_a^b\cos{|z|}+\bar z_az^b\frac{(1-\cos{|z|})}{|z|^2} & i\varepsilon_{acb}z^c\frac{\sin{|z|}}{|z|}\\
-i\varepsilon^{acb}\bar z_c\frac{\sin{|z|}}{|z|} & \delta^a_b\cos{|z|}+z^a\bar z_b\frac{(1-\cos{|z|})}{|z|^2}
\end{array}\right)
\end{equation}
one finds the following expansion
\begin{equation}
\begin{array}{rl}
T_{a}{}^{b}=&T^b{}_a=\left(1-\frac12|z|^2+\frac{1}{24}|z|^4\right)\delta_a^b+\frac12\bar z_az^b\left(1-\frac{1}{12}|z|^2\right)+O(6),\\[0.2cm]
T_{ab}=&i\varepsilon_{acb}z^c(1-\frac16|z|^2)+O(5),\\[0,2cm]
T^{ab}=&-i\varepsilon^{acb}\bar z_c(1-\frac16|z|^2)+O(5).
\end{array}
\end{equation}
Analogously the $\mathbb{CP}^3$ complex vielbein and bosonic part
of the RR 1-form that for the chosen $\mathbb{CP}^3$
parametrization read \cite{U}
\begin{equation}
\begin{array}{c}
\Omega_{a}(d)=d\bar z_a\frac{\sin{|z|}}{|z|}+\bar z_a\frac{\sin{|z|}(1-\cos{|z|})}{2|z|^3}(dz^c\bar z_c-z^cd\bar z_c)+\bar z_a\left(\frac{1}{|z|}-\frac{\sin{|z|}}{|z|^2}\right)d|z|,\\[0.2cm]
\Omega^a(d)=dz^a\frac{\sin{|z|}}{|z|}+z^a\frac{\sin{|z|}(1-\cos{|z|})}{2|z|^3}(z^cd\bar
z_c-dz^c\bar
z_c)+z^a\left(\frac{1}{|z|}-\frac{\sin{|z|}}{|z|^2}\right)d|z|,
\end{array}
\end{equation}
\begin{equation}
\tilde\Omega_a{}^a(d)=i\textstyle{\frac{\sin^2{|z|}}{|z|^2}}(dz^a\bar z_a-z^ad\bar z_a)
\end{equation}
admit the series expansions
\begin{equation}
\begin{array}{c}
\Omega_a(d)=\Omega_{a,b}dz^b+\Omega_{a}{}^{,b}d\bar z_b:\quad
\Omega_{a,b}=\bar z_a\bar z_b(\frac14+\frac16|z|-\frac{1}{16}|z|^2)+O(6),\\[0.2cm]
\Omega_{a}{}^{,b}=(1-\frac16|z|^2+\frac{1}{120}|z|^4)\delta_a^b-\bar z_az^b(\frac14-\frac16|z|-\frac{1}{16}|z|^2)+O(6);\\[0.2cm]
\Omega^a(d)=\Omega^{a}{}_{,b}dz^b+\Omega^{a,b}d\bar z_b:\quad \Omega^{a}{}_{,b}=\Omega_{b}{}^{,a},\\[0.2cm]
\Omega^{a,b}=z^az^b(\frac14+\frac16|z|-\frac{1}{16}|z|^2)+O(6);\\[0.2cm]
\tilde\Omega_a{}^a(d)=\tilde\Omega_a{}^{a}{}_{,b}dz^b+\tilde\Omega_a{}^{a,b}d\bar z_b:\\[0.2cm]
\tilde\Omega_a{}^{a}{}_{,b}=\frac{i}{2}\bar
z_b(1-\frac13|z|^2)+O(5),\
\tilde\Omega_a{}^{a,b}=-\frac{i}{2}z^b(1-\frac13|z|^2)+O(5).
\end{array}
\end{equation}

Taking into account the above expansions the quadratic light-cone gauge Hamiltonian is found to be
\begin{equation}
\mathscr H^{(2)}_{l.c.}=\mathscr H^{(2)}_b+\mathscr H^{(2)}_f
\end{equation}
with the bosonic and fermionic contributions given by
\begin{equation}
\mathscr H^{(2)}_b=\frac12(p^2_1+\textstyle{\frac{\tilde
T^2}{16}}\partial_\sigma x^1\partial_\sigma
x^1)+\textstyle{\frac18}(p^2_\varphi+\tilde
T^2\partial_\sigma\varphi\partial_\sigma\varphi)+\frac12(p_a\bar
p^a+\textstyle{\frac{\tilde T^2}{4}}\partial_\sigma
z^a\partial_\sigma\bar z_a)
\end{equation}
and
\begin{equation}
\mathscr H^{(2)}_{f}=\textstyle{\frac{\tilde
T}{2}}(\eta_a\partial_\sigma\bar\theta^a
+\partial_\sigma\theta_a\bar\eta^a)+\textstyle{\frac{\tilde
T}{4}}(\partial_\sigma\theta_4\bar\eta^4-\partial_\sigma\eta_4\bar\theta^4+
\eta_4\partial_\sigma\bar\theta^4-\theta_4\partial_\sigma\bar\eta^4).
\end{equation}
Corresponding quadratic light-cone gauge Lagrangian can be written in the following form after integrating out the momenta
\begin{equation}\label{l2lc}
\begin{array}{rl}
\mathscr L^{(2)}_{l.c.}=&\frac{1}{2}(\partial_\tau x^1\partial_\tau x^1-\frac{\tilde T^2}{16}\partial_\sigma x^1\partial_\sigma x^1)
+2(\partial_\tau\varphi\partial_\tau \varphi-\frac{\tilde T^2}{16}\partial_\sigma\varphi\partial_\sigma\varphi)\\[0.2cm]
+&2(\partial_\tau z^a\partial_\tau\bar z_a-\frac{\tilde T^2}{16}\partial_\sigma z^a\partial_\sigma\bar z_a)\\[0.2cm]
-&i(\partial_\tau\theta_A\bar\theta^A-\theta_A\partial_\tau\bar\theta^A
+\partial_\tau\eta_A\bar\eta^A-\eta_A\partial_\tau\bar\eta^A)-\frac{\tilde
T}{2}(\eta_A\partial_\sigma\bar\theta^A
+\partial_\sigma\theta_A\bar\eta^A).
\end{array}
\end{equation}
Its bosonic part is readily observed to coincide with that of the flat-space string Lagrangian in the light-cone gauge.
To bring the fermionic terms to the form exhibiting the $2d$ Dirac Lagrangian structure introduce 8-component spinor fields
\begin{equation}
\begin{array}{c}
\Psi_{A'}(\tau,\sigma)=\theta_{A'}-\gamma^7_{A'B'}\bar\eta^{B'},\ \bar\Psi^{A'}=(\Psi_{A'})^\dagger;\quad
\Phi_{A'}(\tau,\sigma)=\theta_{A'}+\gamma^7_{A'B'}\bar\eta^{B'},\ \bar\Phi^{A'}=(\Phi_{A'})^\dagger,
\end{array}
\end{equation}
where
\begin{equation}
\theta_{A'}=\left(
\begin{array}{c}
\theta_A\\
\bar\theta^A\\
\end{array}\right),\quad
\eta_{A'}=
\left(
\begin{array}{c}
\eta_A\\
\bar\eta^A\\
\end{array}\right)
\end{equation}
and
\begin{equation}
\gamma^7_{A'B'}=i\left(
\begin{array}{cc} 0 & -\delta_A^B\\
\delta^A_B & 0
\end{array}\right)
\end{equation}
is one of the $8\times8$ $D=7$ $\gamma-$matrices (see Appendix A
of \cite{me-lc} for the spinor algebra). Introduced spinor fields
satisfy the $7d$ Majorana condition
$\bar\Psi^{A'}=-C^{A'B'}\Psi_{B'}$ etc. (and in fact are the $8d$
Majorana-Weyl spinors of the opposite chirality). As a result the
fermionic part of the Lagrangian (\ref{l2lc}) acquires the form
\begin{equation}\label{fquad}
\begin{array}{c}
-\frac{i}{2}\bar\Psi^{A'}\partial_{-}\Psi_{A'}-\frac{i}{2}\bar\Phi^{A'}\partial_{+}\Phi_{A'},
\end{array}
\end{equation}
where the world-sheet chiral Dirac operators are
$\partial_{\pm}\equiv\partial_\tau\pm\frac{\tilde
T}{4}\partial_\sigma$. Thus the final form of the quadratic
Lagrangian
\begin{equation}
\begin{array}{c}
\mathscr L^{(2)}_{l.c.}=\frac{1}{2}(\partial_\tau x^1\partial_\tau x^1-\frac{\tilde T^2}{16}\partial_\sigma x^1\partial_\sigma x^1)
+2(\partial_\tau\varphi\partial_\tau \varphi-\frac{\tilde T^2}{16}\partial_\sigma\varphi\partial_\sigma\varphi)\\[0.2cm]
+2(\partial_\tau z^a\partial_\tau\bar z_a-\frac{\tilde
T^2}{16}\partial_\sigma z^a\partial_\sigma\bar
z_a)+\frac{i}{2}\Psi^{A'}\partial_{-}\Psi_{A'}+\frac{i}{2}\Phi^{A'}\partial_{+}\Phi_{A'}
\end{array}
\end{equation}
can be recognized as the flat-space superstring Lagrangian in the light-cone gauge.

Continuing the perturbative expansion of the light-cone gauge Hamiltonian (\ref{hlc}) one derives the following expressions describing the first interaction Hamiltonians
\begin{equation}
{\scriptstyle\sqrt{\tilde P_-}}\mathscr H^{(3)}_{l.c.}=\mathscr H^{(3)}_{b}+\mathscr H^{(3)}_{bf},
\end{equation}
where
\begin{equation}
\mathscr H^{(3)}_{b}=-\textstyle{\frac{\tilde
T^2}{4}}\varphi\partial_\sigma x^1\partial_\sigma
x^1-\frac12\varphi(p^2_\varphi+\tilde
T^2\partial_\sigma\varphi\partial_\sigma\varphi)-2\varphi(p_a\bar
p^a+\textstyle{\frac{\tilde T^2}{4}}\partial_\sigma
z^a\partial_\sigma\bar z_a),
\end{equation}
\begin{equation}
\begin{array}{rl}
\mathscr H^{(3)}_{bf}=&-2\tilde
T\varphi(\eta_a\partial_\sigma\bar\theta^a
+\partial_\sigma\theta_a\bar\eta^a)-i\tilde T(\varepsilon^{abc}\eta_a\bar z_b\partial_\sigma\theta_c+\varepsilon_{abc}\bar\eta^az^b\partial_\sigma\bar\theta^c)\\[0.2cm]
-&\tilde
T\varphi(\partial_\sigma\theta_4\bar\eta^4-\partial_\sigma\eta_4\bar\theta^4+
\eta_4\partial_\sigma\bar\theta^4-\theta_4\partial_\sigma\bar\eta^4)-2(p_a\bar\eta^a\bar\eta^4-\bar p^a\eta_a\eta_4)\\[0.2cm]
+&\varepsilon^{abc}p_a\eta_b\eta_c-\varepsilon_{abc}\bar p^a\bar\eta^b\bar\eta^c+2\tilde T(\eta_a\bar\eta^a+\frac12\Theta)\partial_\sigma x^1\\[0.2cm] 
-&2i\tilde T(\partial_\sigma z^a\eta_a\theta_4+\partial_\sigma\bar z_a\bar\eta^a\bar\theta^4),
\end{array}
\end{equation}
and the quartic Hamiltonian
\begin{equation}
\tilde P_-\mathscr H^{(4)}_{l.c.}=\mathscr H^{(4)}_{b}+\mathscr H^{(4)}_{bf}+\mathscr H^{(4)}_{f},
\end{equation}
where
\begin{equation}
\begin{array}{rl}
\mathscr H^{(4)}_{b}=&\tilde T^2\varphi^2\partial_\sigma x^1\partial_\sigma x^1+\varphi^2(p^2_\varphi+
\tilde T^2\partial_\sigma\varphi\partial_\sigma\varphi)+4\varphi^2(p_a\bar p^a+\frac{\tilde T^2}{4}\partial_\sigma z^a\partial_\sigma\bar z_a)\\[0.2cm]
-&\frac{1}{6}((z^a p_a)^2+(\bar p^a\bar z_a)^2-|z|^2(\bar p^a p_a)-(z^a p_a)(\bar p^b\bar z_b))\\[0.2cm]
+&\frac{\tilde T^2}{24}((\partial_\sigma z^a\bar
z_a)^2+(z^a\partial_\sigma\bar z_a)^2-|z|^2(\partial_\sigma
z^a\partial_\sigma\bar z_a)-(\partial_\sigma z^a\bar
z_a)(z^b\partial_\sigma\bar z_b)),
\end{array}
\end{equation}
\begin{equation}
\begin{array}{rl}
\mathscr H^{(4)}_{bf}=&4\tilde
T\varphi^2(\eta_a\partial_\sigma\bar\theta^a
+\partial_\sigma\theta_a\bar\eta^a)+4i\tilde T\varphi(\varepsilon^{abc}\eta_a\bar z_b\partial_\sigma\theta_c+
\varepsilon_{abc}\bar\eta^az^b\partial_\sigma\bar\theta^c)\\[0.2cm]
-&\tilde T|z|^2(\eta_a\partial_\sigma\bar\theta^a
+\partial_\sigma\theta_a\bar\eta^a)+\tilde T((z^a\eta_a)(\partial_\sigma\bar\theta^b\bar z_b)+(z^a\partial_\sigma\theta_a)(\bar\eta^b\bar z_b))\\[0.2cm]
+&2\tilde
T\varphi^2(\partial_\sigma\theta_4\bar\eta^4-\partial_\sigma\eta_4\bar\theta^4+
\eta_4\partial_\sigma\bar\theta^4-\theta_4\partial_\sigma\bar\eta^4)+8\varphi(p_a\bar\eta^a\bar\eta^4-\bar p^a\eta_a\eta_4)\\[0.2cm]
-&4\varphi(\varepsilon^{abc}p_a\eta_b\eta_c-\varepsilon_{abc}\bar p^a\bar\eta^b\bar\eta^c)+2i(\varepsilon^{abc}p_a\bar z_b\eta_c\bar\eta^4+\varepsilon_{abc}\bar p^az^b\bar\eta^c\eta_4)\\[0.2cm] 
+&i((z^ap_a)-(\bar p^a\bar z_a))\Theta-12\tilde T\varphi(\eta_a\bar\eta^a+\frac12\Theta)\partial_\sigma x^1\\[0.2cm]
-&2i\tilde T(\varepsilon^{abc}\eta_a\bar z_b\eta_c+\varepsilon_{abc}\bar\eta^a z^b\bar\eta^c)\partial_\sigma x^1
+8i\tilde T\varphi(\partial_\sigma z^a\eta_a\theta_4+\partial_\sigma\bar z_a\bar\eta^a\bar\theta^4)\\[0.2cm] 
+&2\tilde T(\varepsilon_{abc}\partial_\sigma z^az^b\bar\eta^c\theta_4\!-\!\varepsilon^{abc}\partial_\sigma\bar z_a\bar z_b\eta_c\bar\theta^4)\!+\!\tilde T((\partial_\sigma z^a\bar z_a)\!-\!(z^a\partial_\sigma\bar z_a))(\theta_4\bar\eta^4\!-\!\eta_4\bar\theta^4),
\end{array}
\end{equation}
\begin{equation}
\mathscr H^{(4)}_f=8\theta_4\bar\theta^4\eta_4\bar\eta^4.
\end{equation}

\section{Conclusion}

We have developed the phase-space formulation for the $AdS_4\times\mathbb{CP}^3$ superstring in the fermionic light-cone gauge of Ref. \cite{me-lc}.
There were imposed bosonic light-cone gauge conditions analogous to those of the phase-space formulation for the flat-space and $AdS_5\times S^5$ string
models \cite{GGRT}, \cite{MTlc}. The resulting light-cone gauge Hamiltonian contains quadratic and quartic terms in the fermionic fields with the highly
non-linear dependence on the transverse bosonic fields interacting with the fermions. Its free (quadratic) part can be cast into the form corresponding
to the flat-space Type IIA superstring in the light-cone gauge.

The nonlinearity of the $AdS_4\times\mathbb{CP}^3$ superstring
Hamiltonian even after the exclusion of the pure gauge degrees of
freedom suggests searching for its reformulation in terms of some
new variables, e.g. those resembling twistors \cite{Penrose}. As
is known the $\mathbb{CP}^3$ manifold is isomorphic to the
projective twistor space and the (super)particle models on the
$AdS-$type backgrounds can be formulated in terms of properly
generalized twistor variables \cite{Zunger}, \cite{BLPS}.

It is as well possible to perform the perturbative quantization in
the simplifying limits, such as the large light-cone
momentum/string tension limit around the $pp-$wave background. For
the nontrivial $pp-$waves of $AdS_5\times S^5$ and
$AdS_4\times\mathbb{CP}^3$ backgrounds this has been done using
the phase-space approach in \cite{Ryzhov}-\cite{FPZ} and
\cite{Astolfi}, \cite{Sundin} respectively. In such a way one is
able to calculate the inverse tension (strong 't Hooft coupling)
corrections to the anomalous dimensions of the dual
Berenstein-Maldacena-Nastase (BMN) operators \cite{BMN}. The
identification of the gauge theory states corresponding to the
flat background and study of the appropriate series expansion
remain the problems for future work.

\section{Acknowledgements}

The author is grateful to A.A.~Zheltukhin for stimulating discussions and the Abdus Salam ICTP, where part of the work was done, for warm
hospitality and support.

\end{document}